\documentclass[useAMS]{mn2e}
\usepackage{graphicx}
\usepackage{lscape}

\def\lesssim{\mathrel{\hbox{\rlap{\hbox{\lower4pt\hbox{$\sim$}}}\hbox{$<$}}}}

\def\gtrsim{\mathrel{\hbox{\rlap{\hbox{\lower4pt\hbox{$\sim$}}}\hbox{$>$}}}}

\def\pg{\mbox{PG~0122+200}}
\def\pp{\mbox{PG~1159$-$035}}

\def\rxj{\mbox{RX~J2117.1+3412}}
\def\v4334{\mbox{V4334 Sgr}}

\title[The internal rotation of \pg]{Probing the internal 
rotation of pre-white dwarf stars with asteroseismology: 
the case of \pg}

\author[A. H. C\'orsico, 
        L. G. Althaus, 
        S. D. Kawaler, 
        M. M. Miller Bertolami, 
        E. Garc\'ia--Berro, \& S. O. Kepler]
       {A. H. C\'orsico$^{1,2}$\thanks{E-mail: acorsico@fcaglp.unlp.edu.ar (AHC)}, 
        L. G. Althaus$^{1,2}$, 
        S. D. Kawaler$^{3}$,  
        M. M. Miller Bertolami$^{1,2}$, \newauthor 
        E. Garc\'ia-Berro$^{4,5}$, and 
        S. O. Kepler$^{6}$\\
        $^{1}$Facultad de Ciencias Astron\'omicas y Geof\'isicas, 
              Universidad Nacional de La Plata, 
              Paseo del Bosque s/n, 
              (1900) La Plata, 
              Argentina\\
        $^{2}$Instituto de Astrof\'isica La Plata (IALP-CONICET)\\
        $^{3}$Department of Physics and Astronomy, 
              Iowa State University,
              12 Physics Hall, Ames, 
              IA 50011, 
              U.S.A.\\
        $^{4}$Departament de F\'\i sica Aplicada, 
              Universitat Polit\`ecnica de Catalunya,
              c/Esteve Terrades 5, 
              08860 Castelldefels, 
              Spain\\
       $^{5}$Institute for Space Studies of Catalonia, 
              c/Gran Capit\`a 2--4, 
              Edif. Nexus 104, 
              08034 Barcelona, 
              Spain\\
       $^{6}$ Departamento de Astronomia, Universidade Federal do 
              Rio Grande do Sul, Av. Bento Goncalves 9500
              Porto Alegre 91501-970, RS, Brazil}
\begin{document}

\date{}

\maketitle

\label{firstpage}

\begin{abstract}
We  put  asteroseismological  constraints  on  the  internal  rotation
profile of  the GW Vir (PG1159-type) star  \pg.  To this end  we employ a
state-of-the-art asteroseismological model for this star and we assess
the  expected  frequency splittings  induced  by  rotation adopting  a
forward  approach  in  which  we  compare  the  theoretical  frequency
separations  with  the  observed  ones  assuming  different  types  of
plausible   internal   rotation   profiles.    We  also   employ   two
asteroseismological  inversion  methods   for  the  inversion  of  the
rotation profile of \pg.   We find evidence for differential rotation
in  this star. We demonstrate  that the  frequency  splittings of  the
rotational  multiplets exhibited by  \pg\  are compatible  with a
rotation profile in  which the central regions are  spinning about
2.4 times faster than the stellar surface.
\end{abstract}


\section{Introduction}

The upper left corner  of the Hertzsprung-Russell diagram is populated
by a  handful of rapidly variable  stars, the so-called  GW Vir stars.
They  are very  hot and  luminous, hydrogen-deficient  pre-white dwarf
stars  characterized by  surface  layers rich  in  helium, carbon  and
oxygen    (Werner   \&   Herwig    2006)   that    exhibit   nonradial
$g$(gravity)-modes with  periods between 5  and 50 min ---  see, e.g.,
Winget \& Kepler (2008)  and Althaus et al.  (2010).  In recent years,
accurate asteroseismology of GW Vir stars has
started to  yield details of  the internal structure  and evolutionary
status  of these  stars.   On  the observational  side,  the works  of
Vauclair et al.   (2002) on \rxj, Fu et al.  (2007)  on \pg, and Costa
et  al.    (2008)  on  \pp\  are  particularly   noteworthy.   On  the
theoretical front, important progress  in the modeling of the internal
structure of PG1159  stars (Althaus et al.  2005;  Miller Bertolami \&
Althaus  2006)  has  made it possible  unprecedented  asteroseismological
inferences for GW Vir stars (C\'orsico et al.  2007ab, 2008, 2009).

Asteroseismology  of GW Vir stars provides  information about the
  stellar  mass,  the  chemical  stratification,  the  luminosity  and
  distance,  and several  other  relevant properties  such as  stellar
  rotation rate and the presence and strength of magnetic fields.  Of
particular interest  in the present investigation is  the potential of
asteroseismology  to   place  constraints  on   stellar  rotation,  an
important  aspect  that  has  been  proved to  be  very  difficult  of
assessing by means of  traditional techniques --- mostly spectroscopy.
Specifically,  rotation removes  the  intrinsic mode  degeneracy of  a
nonradial $g$-mode  characterized by an  harmonic degree $\ell$  and a
radial order $k$.  As a result, each pulsation frequency is split into
multiplets of  $2\ell+1$ frequencies specified by  different values of
the azimuthal index $m$, with $m= 0, \pm 1, \ldots, \pm \ell$ (Unno et
al.  1989).  Rotational splittings in  the power spectrum of a compact
pulsator were first  discovered in the white dwarf  R~548 (Robinson et
al.  1976).  Since then, frequency splittings induced by rotation have
been detected in a number of pulsating white dwarf and pre-white dwarf
stars.  If  the rate of rotation  is slow compared  with the pulsation
frequencies, the  frequency separation  between each component  of the
multiplet is proportional to the  rotation velocity of the star.  This
has enabled  to derive the {\sl  mean} rotation period of  a number of
white  dwarf and  pre-white dwarf  stars.  Interestingly  enough, this
approach  provides rotation  velocities much  more precise  than those
inferred from spectroscopy (Koester et al.  1998; Kawaler 2004).

Going one step forward the simple approach described above, Kawaler et
al.  (1999) were  the first to explore the  potential of the inversion
methods  employed  in  helioseismology  to infer  the  {\sl  internal}
rotation of \pp, the prototype of  GW Vir stars.  They found that \pp\
could be rotating  slightly faster at the center  than at the surface,
but such a small contrast could be also compatible with rigid rotation
within the uncertainties of the observed splittings.  Kawaler  et al. (1999)
also  compared  the  patterns  of  ($m=0$) period  spacings  with  the
rotational splittings of \pp\  and empirically found that the rotation
rate of  this star must  increase with depth.  Recently,  Charpinet et
al.  (2009) have employed a  forward approach aimed at determining the
internal rotation profile of \pp.  They fit the theoretical splittings
corresponding   to  a  model   inspired  in   the  asteroseismological
(non-rotating)  model of  C\'orsico  et al.   (2008)  to the  observed
frequency splittings of \pp.  The forward approach of Charpinet et al.
(2009) revealed  that this star  is rotating  as a  solid body  with a
rotation period of $33.61 \pm 0.59$~h.

In this paper we perform a detailed asteroseismological study aimed at
placing constraints on the internal rotation of \pg, the coolest known
GW  Vir  star ($T_{\rm eff}= 80\,  000 \pm 4\, 000$ K and 
$\log g= 7.5\pm 0.5$; Dreizler \& Heber 1998),  using  the  best  
existing  evolutionary  and  seismic
models.  We present  the observations  and describe  the seismological
model of \pg\ in  Sect. \ref{obser-theori}.  In Sect. \ref{forward} we
explore the internal rotation  of \pg\ employing the forward approach.
In  Sect. \ref{inversion-method}  we perform  rotation  inversions for
\pg\ using the Regularized  Least Square method (Sect. \ref{rls}), and
the function fitting technique (Sect. \ref{ffm}). We conclude in Sect.
\ref{conclusions} by summarizing our findings.

\section{The data and the model}
\label{obser-theori}

\begin{table}
\begin{center}
\caption{Properties of \pg.} 
\begin{tabular}{lcr}
\hline
\hline
Quantity & Spectroscopy & Asteroseismology \\
\hline
$T_{\rm eff}$ [kK]          & $80 \pm 4^{\rm (a)}$      & $81.54_{-1.4}^{+0.8}$     \\
$M_*$ [$M_{\sun}$]          & $0.53\pm 0.1^{\rm (b)}$   & $0.556_{-0.014}^{+0.009}$ \\ 
$\log g$ [cm/s$^2$]         & $7.5 \pm 0.5^{\rm (a)}$   & $7.65_{-0.07}^{+0.02}$    \\ 
$\log (L_*/L_{\sun})$       &  $1.2 \pm 0.5^{\rm (c)}$   & $1.14_{-0.02}^{+0.04}$    \\  
$\log(R_*/R_{\sun})$        &  $-1.67\pm 0.26^{\rm (c)}$ & $-1.73_{-0.01}^{+0.025}$  \\  
$M_{\rm env}$ [$M_{\sun}$]  & ---                       & $0.019 \pm 0.006$         \\  
\hline
\hline
\end{tabular}
\label{table1}
\end{center}
{\footnotesize References:  (a) Dreizler  \& Heber (1998);  (b) Miller
Bertolami \& Althaus (2006); (c) Estimated from $T_{\rm eff}$ and $g$,
along with $M_*$.}
\end{table}

\pg\  currently defines  the locus of  the low-luminosity
red edge of the GW  Vir instability strip.  The photometric variations
of this  star were discovered by  Bond \& Grauer  (1987).  
Some  observational properties of  \pg\ are  summarized in  the second
column   of   Table~\ref{table1}.     We   employ   the   high-quality
observational data on  \pg\ gathered by Fu et  al.  (2007).  They have
presented  multi-site photometric observations  of \pg\  obtained with
the  Whole Earth Telescope  (Nather et  al.  1990)  in 2001  and 2002.
Using   their  data,   together  with   those  obtained   in  previous
observational  runs,  they  succeeded  in  detecting  a  total  of  23
frequencies   corresponding  to   modes  with   $\ell=  1$   and  have
unambiguously  derived  a  mean  period  spacing of  22.9~s.   The  23
frequencies consist  of seven  rotational triplets ($m=  -1,0,+1$) and
two    isolated   frequencies    with   (probably)    $m=    0$.    In
Table~\ref{table2} we  show the seven triplets  extracted from table~5
of Fu et al.  (2007).  The  first three columns show the radial order,
the  harmonic  degree,  and   the  azimuthal  quantum  number  of  the
eigenmodes.  The  fourth and fifth columns give  the pulsation periods
and frequencies,  respectively.  The sixth  column shows the  uncertainty of
the  frequencies.   Finally,  in   the  seventh  column  the  observed
frequency  splittings are listed,  while the  last column  gives their
uncertainties.  The mean value (averaged on the seven triplets) of the
frequency separations is 3.74~$\mu$Hz.  Adopting this value, Fu et al.
(2007) derived a mean rotation period of $37.2$ hours.

\begin{table}
\centering
\caption{The rotational triplets of \pg.}
\begin{tabular}{cccccccc}
\hline
\hline
$k$ & 
$\ell$ & 
$m$ & 
$\Pi_{k \ell m}$ &
$\nu_{k \ell m}$ & 
$\sigma[\nu_{k \ell m}]$ & 
$\delta \nu_{k \ell m}^{\rm O}$ & 
$\sigma[\delta \nu_{k \ell m}^{\rm O}]$ \\
 & 
 & 
 &
$[$s$]$ &
$[\mu$Hz$]$ & 
$[\mu$Hz$]$ & 
$[\mu$Hz$]$ & 
$[\mu$Hz$]$ \\
\hline
12 & 1 &$-1$& 336.28 & 2973.73 & 0.003 &      & \\
   &   &    &        &         &       & 3.58 & 0.029\\
12 & 1 &$0$ & 336.68 & 2970.15 & 0.029 &      & \\
   &   &    &        &         &       & 3.60 & 0.030\\
12 & 1 &$+1$& 337.09 & 2966.55 & 0.007 &      & \\
\hline
14 & 1 &$-1$& 379.55 & 2634.65 & 0.009 &      & \\
   &   &    &        &         &       & 3.76 & 0.013\\
14 & 1 &$0$ & 380.10 & 2630.89 & 0.009 &      & \\
   &   &    &        &         &       & 3.87 & 0.015\\
14 & 1 &$+1$& 380.66 & 2627.02 & 0.012 &        & \\
\hline
15 & 1 &$-1$& 400.41 & 2497.41 & 0.002 &      & \\
   &   &    &        &         &       & 3.59 & 0.004\\
15 & 1 &$0$ & 400.99 & 2493.82 & 0.003 &      & \\
   &   &    &        &         &       & 3.58 & 0.004\\
15 & 1 &$+1$& 401.56 & 2490.24 & 0.003 &      & \\
\hline
17 & 1 &$-1$& 448.79 & 2228.18 & 0.005 &      & \\  
   &   &    &        &         &       & 3.42 & 0.006\\
17 & 1 &$0$ & 449.48 & 2224.76 & 0.004 &      & \\
   &   &    &        &         &       & 3.36 & 0.004\\
17 & 1 &$+1$& 450.16 & 2221.40 & 0.002 &      & \\
\hline
18 & 1 &$-1$& 467.87 & 2137.31 & 0.004 &      & \\
   &   &    &        &         &       & 3.73 & 0.008\\
18 & 1 &$0$ & 468.69 & 2133.58 & 0.007 &      & \\
   &   &    &        &         &       & 3.49 & 0.009\\
18 & 1 &$+1$& 469.46 & 2130.09 & 0.005 &      & \\
\hline
22 & 1 &$-1$& 562.70 & 1777.12 & 0.065 &      & \\
   &   &    &        &         &       & 4.96 & 0.085\\
22 & 1 &$0$ & 564.28 & 1772.16 & 0.056 &      & \\
   &   &    &        &         &       & 4.10 & 0.074\\
22 & 1 &$+1$& 565.59 & 1768.06 & 0.049 &      & \\
\hline
24 & 1 &$-1$& 609.64 & 1640.30 & 0.067 &      & \\
   &   &    &        &         &       & 4.05 & 0.094\\
24 & 1 &$ 0$& 611.15 & 1636.25 & 0.067 &      & \\
   &   &    &        &         &       & 3.28 & 0.078\\   
24 & 1 &$+1$& 612.38 & 1632.97 & 0.040 &      & \\    
\hline
\hline
\end{tabular}
\label{table2}
\end{table}

The theoretical  rotational frequency  splittings of \pg\  were
computed adopting  the non-rotating asteroseismological  model derived
by C\'orsico et  al. (2007b) on the basis of the  modern set of PG1159
fully  evolutionary model  sequences computed  by Miller  Bertolami \&
Althaus (2006).  These models  take into
  account  the complete  evolution  of progenitor  stars, through  the
  thermally  pulsing  asymptotic  giant  branch phase  and  born-again
  episode.  C\'  orsico et  al. (2007b) constrain  the stellar  mass of
  \pg\ by  comparing the observed  period spacing with  the asymptotic
  period spacing and with the average of the computed period spacings.
  Finally,  they employ  the  individual observed  periods  to find  a
  representative seismological  model for \pg.  This
asteroseismological  model  reproduces  the  $m= 0$  observed
periods  (see  Table~\ref{table2})  with  an  average  of  the  period
differences  (theoretical  vs.   observed)  of $\lesssim  0.9$~s,  and
represents  a substantial  improvement  over those  models adopted  in
previous works.  The  characteristics of the asteroseismological model
are    compared    with    those   obtained    spectroscopically    in
Table~\ref{table1}.   As can  be seen,  the  asteroseismological model
accurately reproduces the observational data.

%_____________________________________________________________________

\section{The forward approach: rotational splitting fits}
\label{forward}

Within  this approach  the theoretical  frequency  splittings obtained
varying   the  assumed   rotation  profile   are  compared   with  the
observational ones  until a best  global match is found  (Charpinet et
al.  2009).  The goodness  of the  match between  theoretical ($\delta
\nu_{k  \ell m}^{\rm T}$)  and observed  ($\delta \nu_{k  \ell m}^{\rm
O}$)  rotational  splittings is  described  using  a quality  function
defined as

\begin{equation}
\chi^2=\frac{1}{N_{\rm s}}\sum_{i=1}^{N_{\rm s}}
\frac{1}{\sigma_i^{2}}(\delta\nu_{i}^{\rm T} - 
\delta \nu_{i}^{\rm O})^2,
\label{xi}
\end{equation}

\noindent where we have replaced  the subscripts $(k, \ell, m)$ with a
single integer  index $i$ that  labels the specific splitting  ($i= 1,
\cdots, N_{\rm  s}\equiv 14$). Each term  of the sum  is weighted with
the inverse square of the standard uncertainty ($\sigma_i$) of the observed
splittings, which are derived from the uncertainties in the frequencies given
in  Fu et  al.   (2007) and  are shown  in  the last  column of  Table
\ref{table2}.   This  is at  variance  with  the  preliminar study  of
C\'orsico  \& Althaus  (2010), in  which  the fits  of the  rotational
splitting were  made without weighting the  terms of the  sum, and so,
the impact of the different uncertainties of the observational data on
the final result was neglected. 
%Since these $\sigma_i$ have dimensions
%of frequency,  they need  to be normalized,  and that  is accomplished
%using  the  reciprocal sum  of  $\sigma_i$.  
The  lower the  value  of
$\chi^2$,  the  better  the  match  between the  theoretical  and  the
observed frequency splittings.

The   theoretical  rotational  splittings   are  computed   using  the
expressions resulting  from the perturbative theory to  first order in
$\Omega$  (the rotation  rate) that  assumes that  the  pulsating star
rotates with  a period ($P \equiv  1/\Omega$) much longer  than any of
its pulsation  periods (Unno et  al.  1989).  
Under the  assumption of rigid   rotation  ($\Omega$   constant),  
the   theoretical  frequency splittings are given by:

\begin{equation}
\delta \nu_{k \ell m}^{\rm T}=
-m \Omega \left(1-C_{k \ell} \right), 
\label{rota-rigid}
\end{equation}

\noindent with $m= 0, \pm 1, \ldots, \pm \ell$, and $C_{k \ell}$ being
coefficients that  depend on the eigenfunctions of  the pulsation mode
obtained in the non-rotating  case.  Such coefficients are computed as
(Unno et al. 1989):

\begin{equation}
C_{k  \ell}=\frac{\int_0^{R_*}\rho r^2\left[2\xi_r \xi_t
+\xi_t^2\right]dr}{\int_0^{R_*}\rho r^2\left[\xi_r^2
+\ell(\ell+1)\xi_t^2 \right] dr}
\label{coefi-c1}
\end{equation}

\noindent  where $\xi_r$ and  $\xi_t$ are  the unperturbed  radial and
tangential  eigenfunctions, respectively.  In the  case  of $g$-modes,
when $k$  is large then  $\xi_r \ll \xi_t$,  in such a way  that $C_{k
  \ell} \rightarrow 1 / \ell(\ell+1)$ (Brickhill 1975).

If the  condition of rigid  body rotation is relaxed  and (spherically
symmetric) differential rotation  is assumed, $\Omega= \Omega(r)$, the
frequency splittings are given by (Unno et al. 1989):

\begin{equation}
\delta \nu_{k \ell m}^{\rm T}= -m\ \int_0^{R_*} \Omega(r) K_{k\ell}(r) dr,
\label{rota-diff}
\end{equation}

\noindent  $K_{k\ell}(r)$  being   the  first-order  rotation  kernels
computed from the rotationally  unperturbed eigenfunctions as (Unno et
al. 1989):

\begin{equation}
K_{k \ell}(r)= \frac
{\rho r^2 \left\{\xi_r^2 - 2 \xi_r \xi_t - 
\xi_t^2 \left[\ell(\ell+1) - 1\right] \right \}}
{\int_0^{R_*} \rho r^2 \left[ \xi_r^2 + \ell(\ell+1) \xi_t^2 \right] dr}
\label{rot-kernel}
\end{equation}

\noindent From  Eq. (\ref{rota-diff}) it  is clear that  the frequency
splitting for a given mode is  just a weighted average of the rotation
rate $\Omega(r)$ throughout the  star, being the rotation kernel $K_{k
\ell}(r)$ precisely the weighting function. 

\begin{figure} 
\begin{center}
\includegraphics[clip,width=8.3 cm]{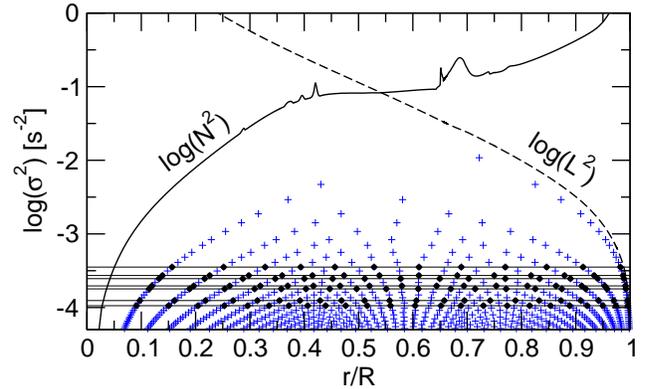} 
\caption{A propagation  diagram of \pg\  showing the logarithm  of the
  squared  Brunt-V\"ais\"al\"a  (solid line)  and  Lamb (dashed  line)
  frequencies. Also  depicted is the  location of the radial  nodes of
  $g$-modes (blue  plus). The loci  of the nodes corresponding  to the
  modes exhibited by \pg\ are emphasized with black dots and connected
  with horizontal lines.} 
\label{propa} 
\end{center}
\end{figure} 

Note that the perturbative theory  to first order in $\Omega$ predicts
symmetric  separations  of  the  $m  \neq 0$  components  within  each
multiplet   with  respect   to   the  central   one   ($m=  0$)   (see
Eqs.  (\ref{rota-rigid}) and  (\ref{rota-diff})).  Therefore,  in this
work  we  are  neglecting  the  departures  from  symmetric  frequency
splitting within  the triplets centered at  $\Pi \sim 560$  s and $\Pi
\sim 610$ s exhibited by \pg\footnote{There exists a number of reasons
  for these departures  (see Vauclair et al. 2002);  for instance, the
  presence of a magnetic field.  The additional frequency shift due to
  a magnetic field  is dependent on $m^2$, and  thus, it could produce
  an asymmetry  in the  frequency shifts  of the $m=  +1$ and  $m= -1$
  components relative to the $m= 0$ component.}.

\begin{figure} 
\begin{center}
\includegraphics[clip,width=8.3 cm]{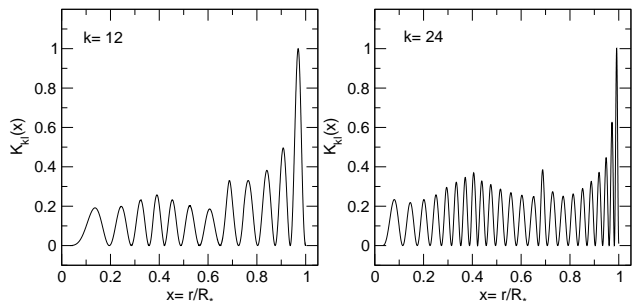} 
\caption{The normalized  rotation kernel $K_{k \ell}(r)$  for the case
  of $k= 12$ (left) and $k= 24$ (right).}
\label{kernel} 
\end{center}
\end{figure}

To  estimate  the  region  of  the star  probed  by  the  observed
$g$-modes,   we  first   examined   a  propagation   diagram  of   the
asteroseismological model of \pg.  In Fig. \ref{propa} we  plot the 
logarithm of the 
squared Brunt-V\"ais\"al\"a  and the Lamb frequencies,  along with the
location of the nodes corresponding to $g$-modes (the zeros of the 
radial eigenfunctions) marked
with  (blue) plus  symbols. The  nodes associated to  the eigenmodes
exhibited by  \pg\ are  emphasized with black  dots.  Note  that these
modes  have nodes in  the region  $0.1 \lesssim r/R_*  \lesssim 1$,
implying that they  have an oscillatory character in  almost the whole
star.   We also  examined  the rotational  kernels  computed from  our
asteroseismological model  for \pg.  In Fig.~\ref{kernel}  we show the
normalized $K_{k\ell}(r)$  for $k= 12$  and $k= 24$,  corresponding to
the shortest  and the longest  pulsation periods observed in  \pg.  As
can be seen,  the rotation kernels have the  largest amplitudes at the
outer regions of  the model, but have also  appreciable amplitudes (up
to $\approx  0.3$) throughout  the {\sl full}  model of  \pg, implying
that  the observed  $g$-modes  are sensitive  to  the entire  rotation
profile. This is in contrast to the case of DBV or DAV stars, in which
rotational kernels sample  only the outer regions of  the star --- see
Kawaler  et al. (1999)  --- because  of the  larger degeneracy  of the
core.

\subsection{Rigid rotation}
\label{rig}

\begin{figure} 
\begin{center}
\includegraphics[clip,width=8.3 cm]{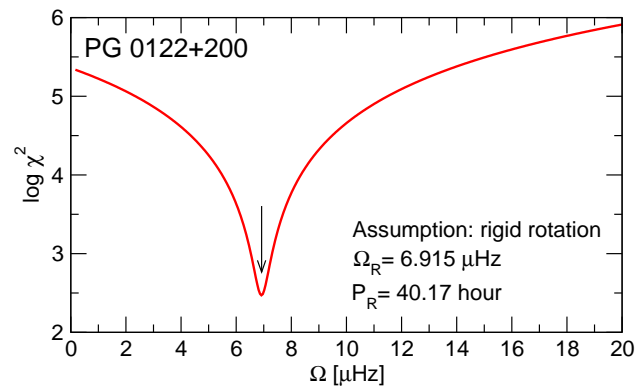} 
\caption{The quality function $\chi^2$ in terms of the 
rotation rate $\Omega$ for the case in which we assume that \pg\
rotates as a rigid body. Note the presence of a 
well defined solution corresponding to $\Omega= 6.915$ $\mu$Hz
(arrow).}
\label{rigid} 
\end{center}
\end{figure}

First, we  assumed that  \pg\ rotates  as a rigid  body, that  is, the
rotation rate  $\Omega$ is constant throughout the  star.  We variated
the  value of  $\Omega$ from  $1.15 \times  10^{-7}$~Hz ($P  \sim 100$
days) to  $2.77 \times 10^{-4}$~Hz ($P  \sim 1$ minutes)  and for each
value  of $\Omega$  we computed  the theoretical  frequency splittings
($\delta \nu_{k \ell m}^{\rm  T}$) by means of Eq.~(\ref{rota-rigid}),
where  the  coefficients $C_{k  \ell}$  were  assessed  for each  mode
through  Eq.~(\ref{coefi-c1})  and  not  by means  of  the  asymptotic
relation, $C_{k \ell} \approx 1  / \ell(\ell+1)$.  The results of this
optimization procedure  are shown in Fig.~\ref{rigid},  that shows the
$\chi^2$  function versus  the rotation  rate.  The  best-fit solution
corresponds  to  a  rotation  rate  of  $\Omega=  6.915$~$\mu$Hz.   It
corresponds to  a rotation  period of $P=  40.17$ hours, in       good
agreement with the  approximate value of $P= 37.2$  hours quoted by Fu
et  al.  (2007).  Note  that  the rotation  period  (of  the order  of
$10^5$~s) is  much longer than the longest  pulsation period exhibited
by \pg\  ($\sim 600$~s), thus  justifying the use of  the perturbative
theory  to a  first order  in  the calculation  of $\delta\nu_{k  \ell
m}^{\rm T}$.

\subsection{Differential rotation}
\label{differential}

\begin{figure} 
\begin{center}
\includegraphics[clip,width=8.3 cm]{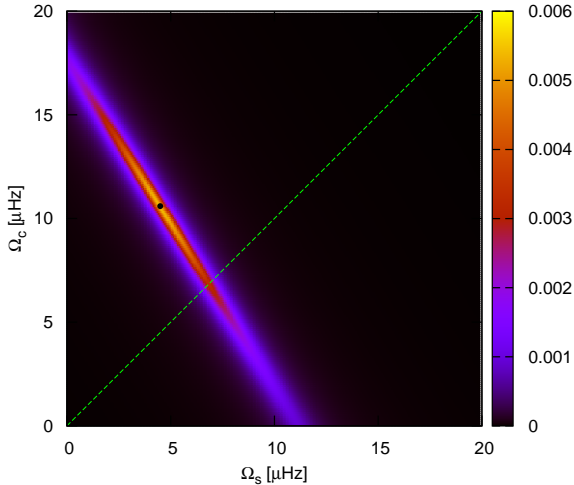} 
\caption{Contour map  of the inverse of the  quality function $\chi^2$ 
         showing  the  goodnesses  of  the fits  in  the  $\Omega_{\rm
         s}-\Omega_{\rm  c}$  plane.   We  assume  that  \pg\  rotates
         according to a linear profile.  The color scale is related to
         the value of $1/\chi^2$.  Light regions are associated to the
         highest values of $1/\chi^2$,  i.e., the best matches between
         observed and theoretical  frequency splittings.  The location
         of the best-fit solution is shown with a black dot. The green
         dashed line indicates the locii of solutions corresponding to
         rigid rotation.}
\label{linear-pg0122} 
\end{center}
\end{figure}

Here,  we lift  the assumption  of solid-body  rotation.   Because the
exploratory nature  of this study,  we try very  simplified functional
forms for $\Omega(r)$. Specifically, we adopt a family of \emph{linear} 
differential rotation profiles defined as:

\begin{equation}
\Omega(r)=(\Omega_{\rm  s}-\Omega_{\rm c}) r + \Omega_{\rm c},  
\label{lineal}
\end{equation}

\noindent where $\Omega_{\rm s}$ and $\Omega_{\rm c}$ are the rotation
rates at the stellar surface and center, respectively\footnote{We have
  also tried  two-zone rotation profiles like those  used by Charpinet
  et  al.  (2009) for  \pp, but  we were  unable to  derive meaningful
  properties  of the  rotation of  \pg,  which is  symptomatic of  the
  inadequacy  of the two-zone  rotation profiles  to represent  the
  internal rotation of \pg.}.  This family of linear profiles includes
rotation rates that  decrease and increase linearly with  $r$ and also
``flat''  rotation profiles  (when $\Omega_{\rm  s}=  \Omega_{\rm c}$)
that  represent  the  case  of  rigid  rotation  already  examined  in
Sect.~\ref{rig}.  We performed our optimization procedures varying the
parameters  $\Omega_{\rm  s}$  and   $\Omega_{\rm  c}$  in  the  range
$0-20$~$\mu$Hz.  We  computed the theoretical  frequency splittings by
means  of  Eq.   (\ref{rota-diff}),  where the  rotation  kernels  are
computed by  using Eq. (\ref{rot-kernel}).   The results are  shown in
Fig.~\ref{linear-pg0122}, where  we prefer to  plot $1/\chi^2$ instead
of  $\chi^2$ to  emphasize the  location of  the  values ($\Omega_{\rm
  c},\Omega_{\rm  s}$) providing good  agreement between  observed and
theoretical frequency  splittings. The region of  good solutions (that
is, the smallest  values of $\chi^2$) has an  elongated shape.  As can
be seen,  there exists a  unique, well-localized best-fit  solution at
$(\Omega_{\rm c},\Omega_{\rm s})=  (10.62, 4.41)\, \mu$Hz, marked with
a black  dot in  the plot.  This  solution is  substantially different
from rigid-rotation, which  should fall at some point  along the green
dashed line.  The existence of the best fit solution suggests that the
central regions of \pg\ could  be rotating more than twice faster than
the surface.

We  studied the  sensitivity of  our result  to each  of  the observed
rotational  triplets.   To  this  end,  we  have  performed  a  simple
experiment:  we  removed  one   of  the  observed  triplets  from  our
computations,  keeping   the  six   remainder,  and  searched   for  a
solution. We started  with the $k= 22$ complex,  that naively could be
thought  to have the  strongest influence  on our  results due  to its
large frequency separations. Indeed, with an average frequency spacing
of $4.53\, \mu$Hz, it is  significantly larger than the other spacings
(see Table~\ref{table2}).   We again find differential rotation with 
the core spinning more  than twice faster
than the surface,  being the variations in the  values of $\Omega_{\rm
s}$ and  $\Omega_{\rm c}$  negligible when compared  with the  case in
which all  the seven  splittings are taken  into account. In  fact, in
this case, we  have $\Omega_{\rm c}= 10.75 \,  \mu$Hz and $\Omega_{\rm
s}=  4.40  \,  \mu$Hz.   THe similarity of the solutions 
is  due  to  its  large  observational
uncertainty, which,  according to Eq.~(\ref{xi}),  strongly attenuates
its  impact.  A  similar experiment  in which  we remove  the  $k= 24$
triplet leads  to very similar results.  Clearly,  these two triplets,
which  suffer from  the largest  uncertainties  in the  list, have  no
appreciable  influence in  our results.   We repeated  this experiment
with  the rest of  the splittings.   We found  that the  most critical
triplets are, in  order of decreasing importance, $k=  18, 17, 14$ and
$15$.   The frequency  separations for  these triplets  are accurately
known.  In particular, if we discard the $k= 18$ triplet, the solution
becomes $\Omega_{\rm c}= 16.45 \,  \mu$Hz and $\Omega_{\rm s}= 0.95 \,
\mu$Hz, that is,  strong differential rotation. On the  other hand, if
the $k= 17$ triplet is not considered, the solution turns out be 
compatible with
rigid rotation, with $\Omega_{\rm  c}= 7.70 \, \mu$Hz and $\Omega_{\rm
  s}= 6.55  \, \mu$Hz.   Hence, our results  rely mostly on  these two
triplets, for which the frequency spacings are well determined 
(although see Sect. \ref{drift}).

Finally, we examined the quality of the match between the observed and
theoretical frequency splittings.  We find that the best fit solution
reproduces the observed frequency splittings with a mean difference of
$\approx 0.25\ \mu$Hz,  which is reduced to $\approx  0.1\ \mu$Hz when
we do not consider the difference corresponding the $k= 22$ triplet.

\begin{figure*} 
\begin{center}
\includegraphics[clip,width=16 cm]{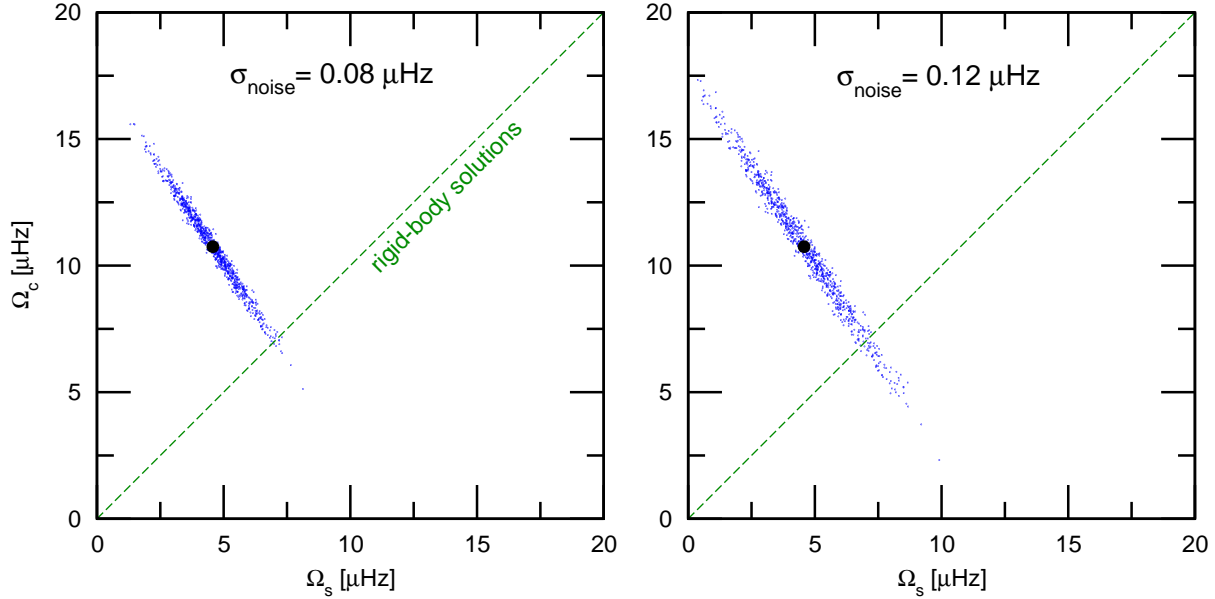} 
\caption{The best-fit solution without uncertainties (black small circle) and
  the  best-fit  solution corresponding  to  400  realizations of  the
  optimization  procedure that  include artificial  noise  (blue dots)
  with  different  standard  deviations ($\sigma_{\rm  noise}$).   The
  green line indicates the locii of solutions of rigid rotation, where
  $(\Omega_{\rm c}=\Omega_{\rm s})$. }
\label{errores} 
\end{center}
\end{figure*} 

\subsection{Uncertainties}
\label{errors}

In order  to assess  the uncertainties in  the derived  parameters, we
repeated  our  optimization   procedure  adding  artificial,  normally
(Gaussian) distributed  uncertainties to the  set of 14  observed splittings,
with a standard deviation of $\sigma_{\rm noise}= 0.08$ $\mu$Hz, which
is comparable with  the best data available for  pulsating white dwarf
and  pre-white dwarf  stars observed  with the  Whole  Earth Telescope
(Kawaler et al.   1999).  We performed about 400  realizations of this
type. The  resulting solutions  are shown in  the left  panel of
Fig.~\ref{errores} with  blue dots. Assuming that  the distribution of
solutions ($\Omega_{\rm  c}$, $\Omega_{\rm  s}$) is also  Gaussian, we
can  estimate its dispersion,  $\sigma$. We  find that  the solutions
that deviate  more from the  best-fit solution (free of uncertainties, black
circle) are located at more than $2 \sigma$ from the line that defines
the solutions  of rigid-body rotation  (green line).  In  other words,
rigid rotation can be discarded at  a level of confidence of more than
$2 \sigma$.  We find  $\Omega_{\rm c}= 10.62  \pm 1.8 \,  \mu$Hz and
$\Omega_{\rm s}= 4.41 \pm 1.1  \, \mu$Hz.  We have repeated the above
analysis adopting  a deviation of $\sigma_{\rm noise}= 0.12 \, \mu$Hz 
for the distribution of uncertainties
in  the frequency  splittings (right panel in
Fig.~\ref{errores}).  These  artificial uncertainties are by  far larger than
the average of  the uncertainties in frequencies quoted by  Fu et al.  (2007)
in their table~4.  Even in this extreme case, we can discard uniform
rotation at  a level of  confidence of more  than $1.5\sigma$.  
Thus,  the conclusion of differential rotation
for  \pg\   remains  unchanged  even  when   we  consider  exaggerated
uncertainties  in  the  measured  frequency splittings.   Finally,  we
investigated the  effects of uncertainties  in the asteroseismological
model on our results, and we found that they are not relevant.

\subsection{The drift of the oscillation frequencies}
\label{drift}

Vauclair et al.  (2011) have  published the results of a comprehensive
monitoring of seven oscillation frequencies of this star.  They report
changes of  these oscillation frequencies over time,  with much larger
amplitudes  and shorter time  scales than  those expected  by cooling,
although  the data  resolution is  rather low,  of the  order  of $1\,
\mu$Hz.  We  focus on Table 4  of Vauclair et al.  (2011), which shows
the  frequency and  amplitude variations  of the  7  largest amplitude
modes of \pg\ corresponding to  the triplets centered at $2224\ \mu$Hz
($k= 17$)  and $2493\  \mu$Hz ($k= 15$).   These two triplets  are the
only ones (out of seven triplets  present in the star) which have been
well documented to exhibit changes over time. 

From Table  4 of Vauclair et  al. (2011) we have  computed the average
value of the rotational  shifts for each triplet as: 

$$
\overline{\delta \nu}(t_n)= \overline{\delta \nu}_n=  
\frac{\delta \nu_{(+)}+  \delta \nu_{(-)}}{2},  
$$

\noindent where $\delta \nu_{(+)}= \nu_{(m= +1)}-\nu_{(m= 0)}$ and 
$\delta \nu_{(-)}= \nu_{(m=   0)}-\nu_{(m= -1)}$ at a given  epoch 
$t_n$ of observation, where $n=
1,\cdots,  N_{\rm obs}$, being $N_{\rm obs}= 6$  for the  triplet  
centered  at  $2224\ \mu$Hz  and  $N_{\rm obs}= 8$ for the triplet  
centered at $2493\ \mu$Hz.  In this way,
we  are discarding  the effects  of  the observed  asymmetries in  the
frequency splittings  within both triplets of \pg.  This is consistent
with the  fact that, in this  work, we are using  the the perturbative
theory  to  first  order   in  $\Omega$  for  estimate  the  frequency
splittings, that  indeed does not  account for possible  departures of
uniformity  of  the splittings  within  a  given  multiplet 
(see Sect. \ref{forward}). Next,  we
estimated an average value of  the splittings over the complete set of
observations, namely 

$$
\langle \overline{\delta  \nu}\rangle= \frac{1}{N_{\rm obs}} 
\sum_{n=1}^{N_{\rm obs}} \overline{\delta   \nu}_n,
$$  

\noindent and the fluctuations around this value 
$\Delta_n=\overline{\delta \nu}_n- \langle \overline{\delta  \nu} \rangle$ 
($n= 1,\cdots,  N_{\rm obs}$). Finally, we computed the  mean value  
of these  fluctuations around  the average  value, 

$$
\langle \Delta_n \rangle= \frac{1}
{N_{\rm obs}} \sum_{n= 1}^{N_{\rm obs}} |\Delta_n|
$$

\noindent and the variance of the fluctuations

$$
\sigma_{\Delta_n}= \sqrt{\frac{1}{N_{\rm obs}} \sum_{n=1}^{N_{\rm obs}}
\left(\Delta_n \right)^2}
$$

\noindent We find for the two triplets $\langle \Delta_n \rangle= 
0.106-0.113\  \mu$Hz and $\sigma_{\Delta_n}= 0.05-0.06\ \mu$Hz.  

We  check  our results  of  non-rigid  rotation  by considering  these
possible variations  of the rotational splittings.   We have performed
new simulations of our optimization  procedure, in each of them adding
Gaussian noise  to the splittings  of the seven triplets  exhibited by
the  star, with  a  standard deviation  $\sigma_{\rm  noise}$.  To  be
consistent,   we   should   adopt   $   \sigma_{\rm   noise}   \approx
\sigma_{\Delta_n}  \sim  0.055\  \mu$Hz   in  order  to  estimate  the
uncertainties  of the frequency  splittings due  to the  observed time
variations,  but  we prefer  to  be  somewhat  conservative and  adopt
$\sigma_{\rm   noise}\approx   \langle \Delta_n \rangle\sim   0.11\,  
\mu$Hz,   thus
overestimating to a some extent  the impact of the frequency drifts on
our results\footnote{Note  that this is actually an  estimation of the
  magnitude of  time variations of  the frequency separations  of \pg,
  because we  do not know the  changes that could  be experiencing the
  frequencies and  frequency separations that have  not been monitored
  by  Vauclair et  al.  (2011).}.   As  expected, the  results of  our
simulations are very similar to those shown in the right panel of Fig.
\ref{errores}, indicating that rigid body rotation can be discarded at
a level of confidence of more than $\sim 1.5 \sigma$.

%_____________________________________________________________________

\section{The inverse problem}
\label{inversion-method}

\subsection{The RLS inversion fits}
\label{rls}

We  also investigated  the internal  rotation rate  of \pg\  using the
Regularized  Least   Squares  (RLS)  fitting   technique  (Kawaler  et
al. 1999) that has been extensively applied to the case of the Sun ---
see Christensen-Dalsgaard  et al.  (1990) and  references therein.  In
this  method, the  internal rotation  profile $\Omega(r)$  is obtained
by inverting the equation (Jeffrey 1988):

\begin{equation}
\delta \nu_{i}^{\rm O}= \int_0^{R_*} \Omega(r) K_{i}(r) dr 
\label{integral}
\end{equation}

\noindent  where  $\delta  \nu_{i}^{\rm  O}$  is  the  $i$th  observed
frequency separation  ($i= 1, \cdots, M$,  where $M$ is  the number of
observed  multiplets). Here, $\delta  \nu_{i}^{\rm O}$  corresponds to
the average separation within  a given multiplet.  Because the problem
is  intrinsically  ill-posed,  the  inversion of  the  above  equation
necessarily must  be regularized (Jeffrey 1988, Kawaler  et al. 1999).
To do so we minimize:

\begin{equation}
S=\sum_{i=1}^{M} \frac{1}{\sigma_{i}^2}
\left[\delta\nu_{i}^{\rm  O}-  \delta   \nu_{i}^{\rm  T}  \right]^2  +
\lambda \int_0^R \left[{\cal L} \Omega(r)\right]^2 dr.
\end{equation}

\noindent where the second term  of the right hand side is, precisely,
the  regularization  term.   The   form  of  the  regularization  term
determines some additional constraints to the solution.  For instance,
if  $\Omega(r)$ cannot  have  a steep  spatial  gradient, then  ${\cal
L}\equiv d/dr$.   On the  other hand, if  $\Omega(r)$ must  be smooth,
then  ${\cal L}\equiv d^2/dr^2$.   The theoretical  splittings $\delta
\nu_{i}^{\rm T}$ are computed numerically:

\begin{equation}
\delta \nu_{i}^{\rm T}= \int_0^{R_*} \Omega(r) K_{i}(r) dr \approx 
\sum_{j= 1}^{N} w_j K_{ij} \Omega_j
\label{integral2}
\end{equation}

\noindent where $\Omega_j=  \Omega(r_j)$ and $K_{ij}= K_i(r_j)$, being
the index $j$  ($j=1, \cdots, N$) associated to  the radial mesh point
in the  stellar model  on which the  kernel $K_i(r)$ and  the rotation
rate $\Omega(r)$ are evaluated, and $w_j= r_{j+1}-r_j$.

In the least squares method, the  derivative of the sum of the squared
residuals (the function  $S$) with respect to $\Omega_j$  is taken and
equated to zero.  After some algebra, the following matrix equation is
derived:

\begin{equation}
(\mathbf{K}^T \mathbf{K}+
\lambda \mathbf{H}) \mathbf{\Omega}= \mathbf{K}^T \mathbf{\Upsilon},
\label{matricial}
\end{equation}

\noindent where $\mathbf{K}$ is a  $(N \times M)$ matrix with elements
$(\mathbf{K})_{ij}= w_j  K_{ij}/\sigma_i$, $\mathbf{\Omega}$ is  a $(N
\times 1)$  vector with components  $(\mathbf{\Omega})_{j}= \Omega_j$,
and  $\mathbf{\Upsilon}$ is a  $(M \times  1)$ vector  with components
$(\mathbf{\Upsilon})_{i}=  \delta  \nu_i^{\rm O}/\sigma_i$.   Finally,
$\mathbf{H}$ is the $(N \times N)$ regularization matrix, which adopts
a tridiagonal form (that  is, $(\mathbf{H})_{ij}= 0$ for $|i-j|>1$) or
a  pentadiagonal  structure   (that  is,  $(\mathbf{H})_{ij}=  0$  for
$|i-j|>2$)  depending on whether  ${\cal L}\equiv  d/dr$ or  ${\cal L}
\equiv d^2/dr^2$.  Eq.~(\ref{matricial})  constitutes a $(N \times N)$
system  of  linear equations  that  must  be  solved for  the  unknown
rotation velocities $\Omega_j$ that minimize $S$.

We have  tested the  reliability of our  RLS scheme by  employing this
technique on ``synthetic'' (free of uncertainties) frequency splittings 
generated with the asteroseismological model of \pg\ through the
forward approach. Specifically, we considered rotational
splittings corresponding to consecutive 
$\ell= 1$ $g$-modes with $k= 1,\cdots, 40$. 
We  employed  the  LU decomposition and  also the Gauss-Jordan  
methods (Press et  al. 1992)
for solve the system.  Both  methods give almost identical results. In
all of the cases we have  examined, the inversions are able to recover
the  input rotation  profile that  we  used to  compute the  synthetic
splittings, provided that an adequate range of values of the parameter 
$\lambda$ is adopted.

\begin{figure} 
\begin{center}
\includegraphics[clip,width=8.5cm]{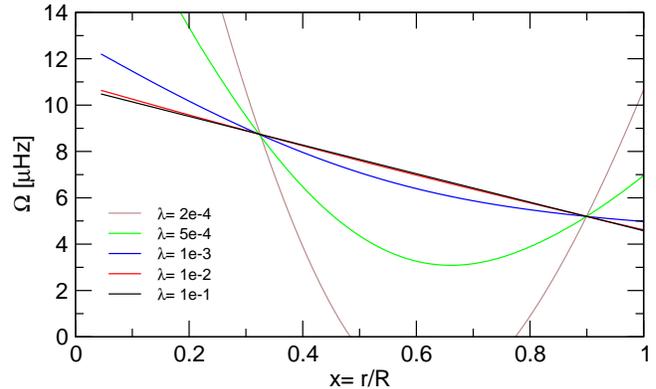} 
\caption{Inverted rotation profiles for \pg\ corresponding to different 
         values of the regularization parameter $\lambda$.}
\label{inv-pg0122} 
\end{center}
\end{figure}

We have applied the RLS  method to infer the internal rotation profile
of \pg.   We have employed the  $M= 7$ averaged  $\ell= 1$ splittings.
The regularization  matrix corresponds to the smoothing  of the second
derivative  of  $\Omega(r)$.   In  Fig.~\ref{inv-pg0122} we  show  the
inverted rotation  profiles for \pg\ for several  values of $\lambda$.
For  very small  values of  $\lambda$, the  inverted  profiles exhibit
strong variations  that lack physical meaning.  However,  as the value
of $\lambda$ is increased, the inverted solution gradually stabilizes.
The  resulting  rotation profile  (corresponding  to $\lambda  \gtrsim
10^{-2}$) consists of an almost linearly decreasing rotation rate with
$\Omega_{\rm c}  \sim 10.75\, \mu$Hz  and $\Omega_{\rm s}  \sim 4.58\,
\mu$Hz,  in  excellent  agreement  with  the results  of  the  forward
approach.   The monotonic  linear functional  form  characterizing the
inverted  rotation profiles  should not  be surprising,  since  we are
forcing $\Omega(r)$ to have a  small value of its second derivative at
the outset.  An  analysis of the uncertainties similar to  that 
performed for the
forward approach leads to the  conclusion that even with the inclusion
of uncertainties in  the observed splittings, the rotation  of \pg\ is
faster at the  central regions than at the  surface.  
Specifically, we found  $\Omega_{\rm c}=  10.75\pm 2.4\,  \mu$Hz 
and  $\Omega_{\rm s}= 4.58\pm 1.7\, \mu$Hz.

\subsection{The function fitting method}
\label{ffm}

We have also made rotational  inversions onto a fixed functional basis
--- a method  called ``function fitting'', see Kawaler  et al.  (1999)
for details. In this inversion technique, an explicit assumption about
the  functional  form  of  $\Omega(r)$  is  made.   For  instance,  if
$\Omega(r)$ is assumed to be a polynomial in $r$ of degree $K-1$, then
$\Omega(r)= \sum_{k=1}^{K} a_k  r^{k-1}$.  Following the least squares
method,  in which we  minimize the  sum of  the squared  residuals, we
obtain the matrix equation:

\begin{equation}
(\mathbf{A}^T \mathbf{A}) \mathbf{a}= \mathbf{A}^T \mathbf{\Upsilon},
\label{matricial2}
\end{equation}

\noindent where $\mathbf{A}$ is a  $(M \times K)$ matrix with elements

\begin{equation}
(\mathbf{A})_{ik}=  \sum_{j= 1}^{N}  \frac{1}{\sigma_i} w_j  K_{ij} r_j^{k-1},
\end{equation}

\noindent  $\mathbf{a}$ is  a $(K  \times 1)$  vector  with components
$(\mathbf{a})_{k}= a_k$,  and $\mathbf{\Upsilon}$ is a  $(M \times 1)$
vector  with  components  $(\mathbf{\Upsilon})_{i}= \delta  \nu_i^{\rm
  O}/\sigma_i$.  Eq.~(\ref{matricial2}) is a  $(K \times K)$ system of
linear equations that must be solved  for the unknown set of values of
$a_k$ that minimize $S$.

Specifically, we  have explored linear  ($K= 2$) functional  forms for
$\Omega(r)$, defined  by the two parameters $a_1=  \Omega_{\rm c}$ and
$a_2= (\Omega_{\rm s}- \Omega_{\rm  c})$.  The optimal values we found
for these parameters are $\Omega_{\rm  c}= 10.74\pm 2.9\, \mu$Hz and
$\Omega_{\rm s}=  4.57\pm 1.8\,  \mu$Hz, in excellent  agreement with
the RLS fits and also with the forward approach.

%_____________________________________________________________________

\section{Summary and conclusions}
\label{conclusions}

Reliable  determinations  of the  rotation  rate  of  white dwarf  and
pre-white  dwarf  stars  are   important,  because  they  can  provide
constraints  on the theories  of angular  momentum transport  from the
core to  the envelopes of  their progenitors.  These  theories predict
white  dwarf   rotation  rates  that   are  much  larger   than  those
spectroscopically found  (Koester et al.  1998; Berger  et al.  2005),
and even larger than the  rotation rates inferred from pulsating white
dwarfs (Kawaler 2004), unless  magnetic torques are taken into account
(Suijs  et al. 2008).   These magnetic  torques provide  the necessary
spin-down of the cores of the white-dwarf progenitors, thus helping to
understand the slow rotation of white dwarfs.

In this work, we have  explored the internal rotation of the pulsating
pre-white  dwarf star  \pg\ on  the  basis of  its observed  frequency
splittings  employing  a  forward   approach  and  also  two  rotation
inversion techniques.  The three methods employed in this work suggest
that the internal  rotation profile of \pg\ is  differential, with the
central regions rotating more  than twice faster than stellar surface.
Rigid  body rotation  can be  discarded for  this star  at a  level of
confidence  of more  than  $1.5$  and $2\,  \sigma$  when we  consider
uncertainties  in the  observed  frequency separations  of $0.12$  and
$0.08\, \mu$Hz, respectively. By averaging our results according
to the  three method  employed, we  estimate that  the core  to surface
  rotation ratio is $\Omega_{\rm c}/\Omega_{\rm s}= 2.4 \pm 1.3$.

Up to now,  only our Sun (Schou et al. 1998)
and  two pulsating  $\beta$  Cephei  stars ---  HD  129929 (Dupret  et
al. 2004) and $\nu$ Eridani (Pamyatnykh  et al. 2004) --- are known to
be rotating  differentially with depth.   Our results for \pg\  can be
considered  as  the   first asteroseismic  evidence  of  differential
rotation with depth  for an evolved star.
Finally, the fact that  the internal rotation profile of  the 
prototypical star
\pp, which  is rotating as a  rigid body, is
apparently very  different than that  obtained here for \pg\  could be
indicative that PG1159 stars  are the result of different evolutionary
channels  and different  mass-loss histories,  as suggested  by recent
theoretical evidence (Althaus et al.  2009).  Our results call for the
need of similar analysis for other well studied GW Vir stars.

\section*{Acknowledgments}

Part of  this work  was supported by  AGENCIA through the  Programa de
Modernizaci\'on   Tecnol\'ogica    BID   1728/OC-AR,   by    the   PIP
112-200801-00940    grant    from     CONICET,    by    MCINN    grant
AYA2008--04211--C02--01,  by  the  ESF EUROCORES  Program  EuroGENESIS
(MICINN grant  EUI2009-04170), by the European Union  FEDER funds, and
by the AGAUR.  This research  has made use of NASA's Astrophysics Data
System.

\label{lastpage}


\begin{thebibliography}{99}

\bibitem{} Althaus, L. G., C\'orsico A. H., Garc\'ia-Berro, E,
           \& Isern, J. 2010, A\&AR, 18, 471
\bibitem{} Althaus, L. G., Panei, J. A., Miller Bertolami, M. M., 
           Garc{\'{\i}}a-Berro,  E., C\'orsico,  A.  H., Romero,  
           A. D.,  Kepler, S. O., \& Rohrmann, R. D., 2009, ApJ, 
           704, 1605
\bibitem{} Althaus, L. G., Serenelli, A. M., Panei, J. A., et al., 
           2005, A\&A, 435, 631
\bibitem{} Berger, L., Koester, D., Napiwotzki, R., Reid, I. N., \& 
           Zuckerman, B., 2005, A\&A, 444, 565
\bibitem{} Brickhill, A. J. 1975, MNRAS, 170, 405 
\bibitem{} Charpinet, S., Fontaine, G., \& Brassard, P., 2009, Nature, 
           461, 501
\bibitem{} Christensen-Dalsgaard, J., Schou, J., \& Thompson, M. J.,
           1990, MNRAS, 242, 353
\bibitem{} C\'orsico, A. H., \& Althaus, L. G.  2010, 
           American Institute of Physics Conference Series, 1273, 566 
\bibitem{} C\'orsico, A. H., Althaus, L. G., Miller Bertolami, M. M., 
           \& Garc\'ia-Berro, E., 2009, A\&A, 499, 257
\bibitem{} C\'orsico, A. H., Althaus, L. G., Kepler, S. O., Costa, J. E. S., 
           \& Miller Bertolami, M. M., 2008, A\&A, 478, 869
\bibitem{} C\'orsico, A. H., Althaus, L. G., Miller Bertolami, M. M., 
           \& Werner, K., 2007a, A\&A, 461, 1095
\bibitem{} C\'orsico, A. H., Miller Bertolami, M. M., Althaus, L. G., 
           Vauclair, G., \& Werner, K., 2007b, A\&A, 475, 619
\bibitem{} Costa, J. E. S., Kepler, S. O., Winget, D. E. et al. 2008, 
           A\&A, 477, 627
\bibitem{} Dreizler, S., \& Heber, U., 1998, A\&A, 334, 618
\bibitem{} Dupret, M. -A, Thoul, A., Scuflaire, R., 
Daszy\'nska-Daszkiewicz, J., Aerts, C., Bourge, P.-O., Waelkens, C., 
Noels, A., 2004, A\&A, 415, 251 
\bibitem{} Fu, J.-N., Vauclair, G., Solheim, J.-E., et al., 2007, 
           A\&A, 467, 237
\bibitem{} Jeffrey, W., 1988, ApJ, 327, 987
\bibitem{} Kawaler, S. D., 2004, in {\sl ``Stellar Rotation''}, Ed. A. 
           Maeder, \& P. Eenens, ASP, IAU Symp., 215, 561
\bibitem{} Kawaler, S. D., Sekii, T., \& Gough, D., 1999, ApJ, 516, 349
\bibitem{} Koester, D., Dreizler, S., Weidemann, V., \& Allard, N. F., 1998, 
           A\&A, 338, 612
\bibitem{} Miller Bertolami, M. M., \& Althaus, L. G., 2006, A\&A, 454, 845 
\bibitem{} Nather, R. E., Winget, D. E., Clemens, J. C., Hansen, C. J., 
           \& Hine, B. P., 1990, ApJ, 361, 309
\bibitem{} Pamyatnykh, A. A., Handler, G., Dziembowski, W. A., 2004, 
MNRAS, 350, 1022

\bibitem{} Press, W. H., Teukolsky, S. A., Vetterling, W. T., \& Flannery, 
B. P. 1992, Cambridge: University Press, 2nd ed.

\bibitem{} Robinson, E. L., Nather, R. E., \& McGraw, J. T., 1976, 
           ApJ, 210, 211
\bibitem{} Schou, J., Antia, H. M., Basu, S., et al. 1998, ApJ, 505, 390 

\bibitem{} Suijs, M. P. L., Langer, N., Poelarends, A.-J., Yoon, S.-C., 
           Heger, A., \& Herwig, F. 2008, A\&A, 481, L87
\bibitem{} Unno, W., Osaki, Y., Ando, H., Saio, H., \& Shibahashi, H.,  
           1989, {\sl ``Nonradial Oscillations  of  Stars''}, University  
           of Tokyo  Press, 2nd edition
\bibitem{} Vauclair, G., Fu, J.-N,  Solheim, S. -L, et al., 2011, A\&A, 
           528, A5 
\bibitem{} Vauclair, G., Moskalik, P., Pfeiffer, B., et al., 2002, A\&A, 
           381, 122
\bibitem{} Werner, K., \& Herwig, F., 2006, PASP, 118, 183
\bibitem{} Winget, D. E., \& Kepler, S. O.,  2008, ARAA, 46, 157

\end{thebibliography}
\end{document}